\documentclass[prb,superscriptaddress,twocolumn,preprintnumbers,a4,showpacs,floatfix]{revtex4-1}  
\usepackage{amsmath, amssymb, psfrag, tabls, dcolumn, bm, color}
\usepackage[sort&compress]{natbib}
\usepackage[pdftex]{graphicx}
\usepackage{bm}

\begin{document}

\title{Graphene Cardboard: from Ripples to Tunable Metamaterial}

\author{Pekka Koskinen}
\email[email:]{pekka.koskinen@iki.fi}
\address{NanoScience Center, Department of Physics, University of Jyv\"askyl\"a, 40014 Jyv\"askyl\"a, Finland}




\begin{abstract}
Recently graphene was introduced with tunable ripple texturing, a nanofabric enabled by graphene's remarkable elastic properties. However, one can further envision sandwiching the ripples, thus constructing composite nanomaterial, graphene cardboard. Here the basic mechanical properties of such structures are investigated computationally. It turns out that graphene cardboard is highly tunable material, for its elastic figures of merit vary orders of magnitude, with Poisson ratio tunable from $10$ to $-0.5$ as one example. These trends set a foundation to guide the design and usage of metamaterials made of rippled van der Waals solids. 
\end{abstract}

\maketitle

Carbon nanomaterials have huge variations in their mechanical and electronic properties. Materials such as graphene, diamond-like carbon, carbon aerogels, and soot show widely varying properties regarding porosity, surface area, electric conductivity, chemical reactivity, and optical and mechanical properties.\cite{morris_2013,mintmire_PRL_92} However, certain nanomaterials can also be seen as building blocks for more complex nanomaterials. A timely example are nanotube forests, which were grown to provide elastic raw material for twist-spun nanotube yarns used as custom-made artificial muscles.\cite{lima_science_12} While efforts to build composite materials have often been exploiting carbon nanotubes, also graphene is equally elastic and serviceable as a raw material for more complex nanomaterials.\cite{kudin_PRB_01, Geim2007,CastroNeto2009}



\begin{figure}
\includegraphics[width=8cm]{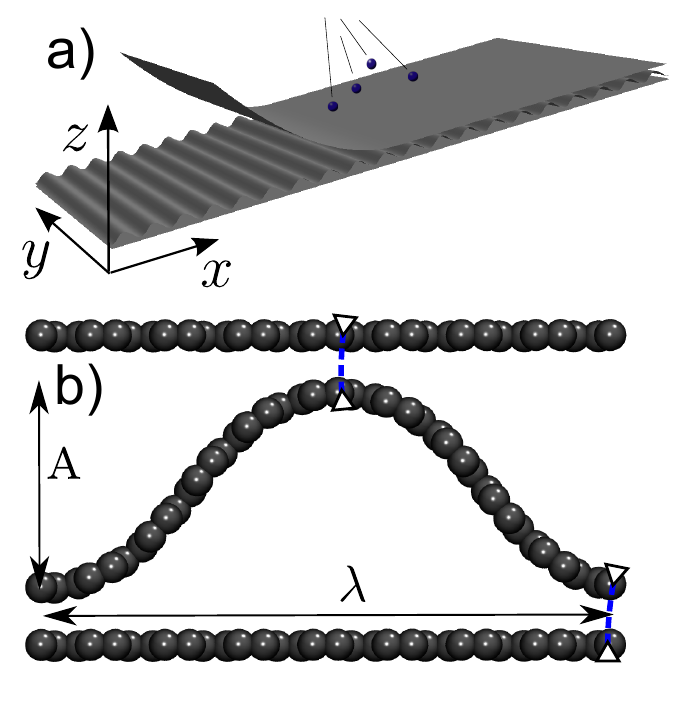}
\caption{Graphene cardboard. (a) A schematic of a rippled graphene membrane sandwitched between two or more graphene layers and welded together by electron irradiation. (b) The basic geometry of a ripple of length $\lambda$ and amplitude $A$, attached by constrained distances to the layers above and below.}
\label{fig:concept}
\end{figure}

Recent experiments reported free-standing periodic rippling in graphene, analogously to the rippling of satin sheets under shear.\cite{Bao2009} The ripples had tunable wavelengths with $\lambda=0.37\ldots 5$ $\mu$m and amplitudes with $A=7\ldots 30$~\AA. Further experiments reported similar findings, including much smaller ripples with $\lambda=7$~\AA\ and $A=0.5$~\AA.\cite{Chen2009,Wang2011a,tapaszto_nphys_12} The ripplings were also supported by theory.\cite{Duan2011} However, it is easy to envision sandwitching the rippled graphene, to construct a single composite nanostructure, graphene cardboard of sort (Fig.\,\ref{fig:concept}a). This would be a way of using graphene as a building block to construct customized composite nanomaterials.


In this Letter, by using theoretical continuum modeling, I investigate the main structural and mechanical characteristics of graphene cardboards of various kind. The aim is to explore structural phase diagram and to calculate the elastic figures of merit. As it turns out, graphene cardboards are highly tunable materials, displaying positive and negative Poisson ratios and large variations in elastic moduli. The results are intuitive and easily generalizable to other layered van der Waals materials.


To model the cardboard, a ripple of wavelength $\lambda$ and amplitude $A$ is sandwitched between flat layers of graphene above and below (Fig.\,\ref{fig:concept}b). The ripple's crests have a constrained distance $h=3.4$~\AA\ to the sandwitch layers. This constraint is applied for clarity, and the actual distance had minor effect on the key results. In practice the distance would be determined by patchy covalent bonds holding the sandwitch together, introduced either by electron irradiation or by chemical functionalization.\cite{Krasheninnikov2007,Locatelli2013} Although this is a schematic model, it well suffices to capture the essentials of the cardboard geometry. The parameters spanning the structural phase space are the wavelength $\lambda$ and the shrinkage of the rippled layer in $x$-direction, the apparent "strain" $\varepsilon_x=(l-\lambda)/l$, where $l$ is the unstrained length prior to rippling. The amplitude $A$ is an outcome of structural optimization; deformations are introduced by constraining the amplitude or crests' lateral displacements.

\begin{figure}
\includegraphics[width=8cm]{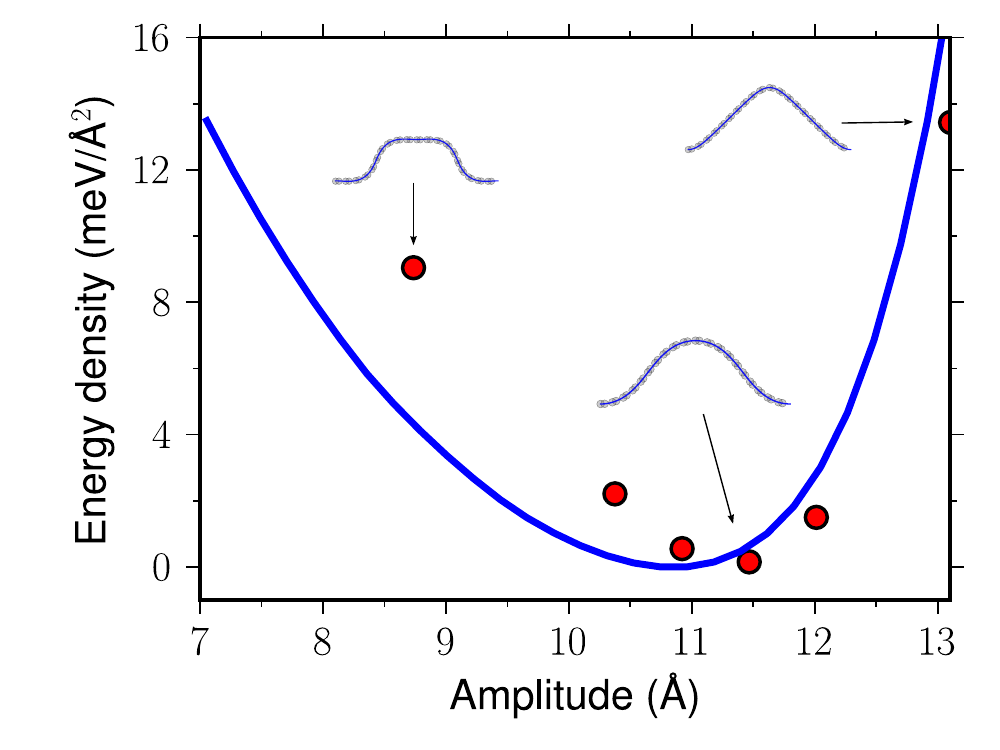}
\caption{Comparison of total energy from DFT (circles) and elastic model (curve). Insets show the superimposed ripple profiles from DFT (circles) and from elastic model (curves). The energy minima are set to zero.}
\label{fig:dft}
\end{figure}

The initial modeling was done by density-functional theory (DFT) simulations that used a van der Waals (vdW) xc-functional (see Supplemental Material \cite{supplemental}). With $\lambda=34.1$~\AA\ and $\varepsilon_x=0.2$ the structural optimization resulted in the familiar sine-type ripple with an amplitude $A=11.4$~\AA\ (Fig.\,\ref{fig:dft}). Compression resulted in a square-shaped profile and tension in a volcano-shaped profile. These results are not surprising, as graphene has repeatedly been reported to behave like a classical elastic membrane despite its atomic thickness.\cite{Lier2000,kudin_PRB_01,topsakal_PRB_10,kit_PRB_12} Also here, therefore, I compare results from DFT to results from an elastic model containing bending energy (proportional to bending modulus $\kappa$), strain energy (proportional to in-plane modulus $k$), and van der Waals energy (proportional to adhesion energy $\varepsilon_\text{vdW}$); see Supplemental Material for details.\cite{supplemental} As anticipated, quantum and classical models agree fairly well regarding ripple profiles, deformation energetics, and optimal amplitudes (Fig.\,\ref{fig:dft}). Calculations with other values of $\lambda$ and $\varepsilon_x$ resulted in a similar agreement. The agreement is not perfect, but it well suffices for the trend-hunting purposes of this work. Thus, the elastic model was the method of choice to proceed calculating material properties as a function of $\lambda$ and $\varepsilon_x$.


\begin{figure}[t!]
\includegraphics[width=8cm]{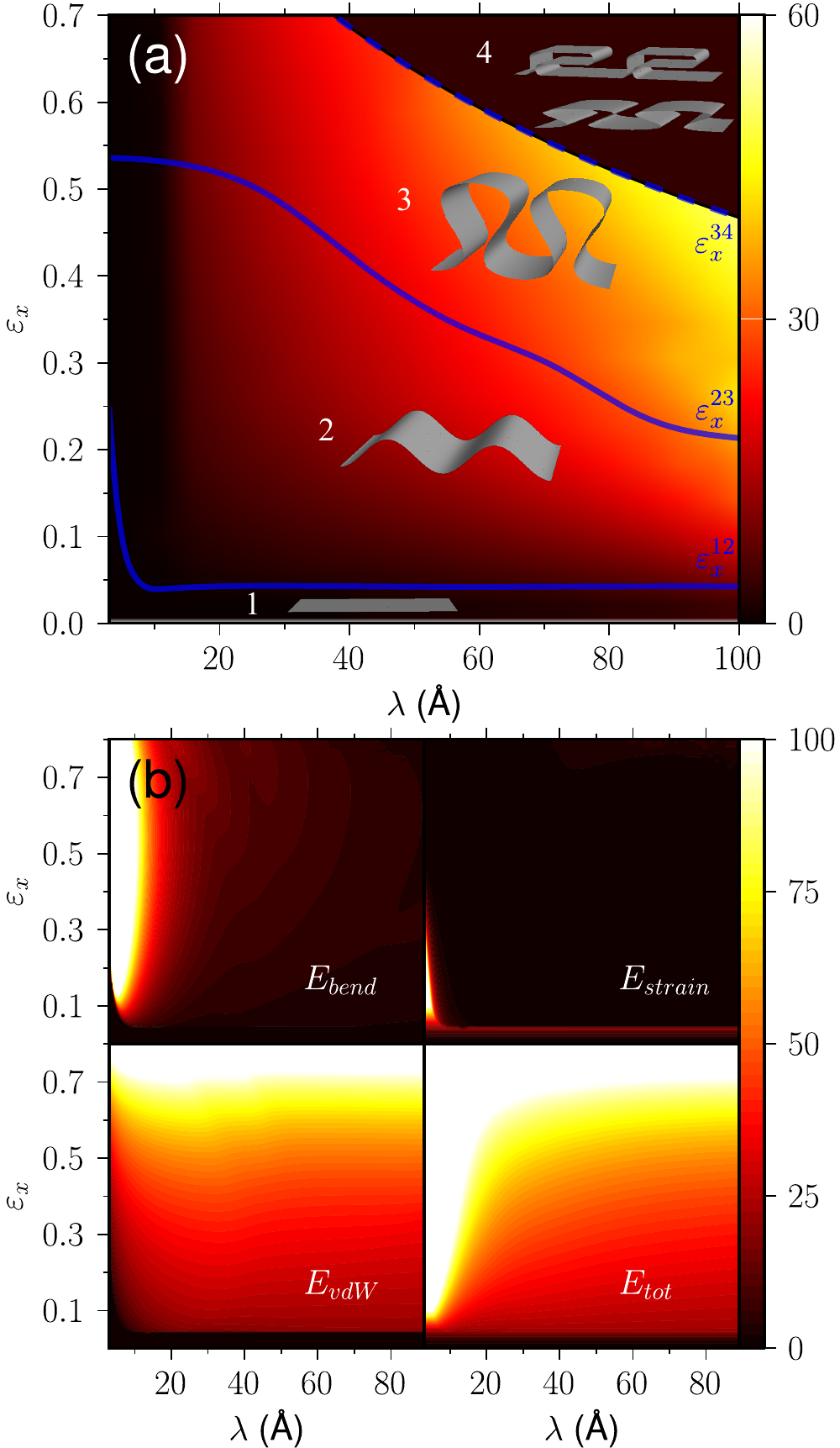}
\caption{Graphene cardboard phase diagram. (a) Contour plot of ripple amplitude $A$ versus ($\lambda,\varepsilon_x$)-plane with four phases: flat layer (1), sine-type ripples (2), mushroom-type ripples (3), and collapsed ripples (4). (b) Areal densities for bending energy ($E_\text{bend}$), strain energy ($E_\text{strain}$), van der Waals energy ($E_\text{vdW}$), and the total energy ($E_\text{tot}$) (for uncollapsed ripples). The minimum of $E_\text{vdW}$ is set to zero and all plots have the same scale with energy maximum of $100$~meV/\AA$^2$.}
\label{fig:phases}
\end{figure}

Optimizing the cardboard geometries for $\lambda=3-100$~\AA\ and $\varepsilon_x=0-0.7$ revealed four phases (Fig.\,\ref{fig:phases}a). In phase $1$, when $\varepsilon_x$ is small, layers remain flat because the compressive stress is yet insufficient to overcome the vdW adhesion. When the apparent strain increases to $\varepsilon_x=\varepsilon_x^{12}(\lambda)$, system makes a transition to a phase $2$ with ripples. At the limit of shallow ripples, the model yields the sine-type profile 
\begin{equation}
z(x)=\left( \frac{2\lambda}{\pi} \sqrt{\frac{\varepsilon_x}{1-\varepsilon_x}} \right) \sin^2 (\pi x/\lambda),
\label{eq:sine}
\end{equation}
where the expression in brackets is the amplitude $A=A(\lambda,\varepsilon_x)$ (see Supplemental Material \cite{supplemental}). At the onset of rippling the amplitude jumps from zero to $A_\text{min}=0.28 h=0.95$~\AA, corresponding to the emergence of two energy minima in the vdW function of the sandwitch [$W_2(z)$ in Eq.\,(1) of Supplemental Material]. The transition $1\rightarrow 2$ occurs when $E_\text{vdW}=15\varepsilon_\text{vdW}A^2/h^2$ at $A=A_\text{min}$ becomes less than $E_\text{strain}=k\varepsilon_x^2/2$, yielding $\varepsilon_x^{12}=1.5\sqrt{\varepsilon_\text{vdW}/k}=0.044$, in agreement with the simulations. This estimate however applies only for $\lambda \gtrsim \lambda_{co}=10$~\AA, a length scale discussed more later. Then, as $\varepsilon_x$ increases further, ripple grows and begins deviating from the sine-profile, until at $\varepsilon_x=\varepsilon_x^{23}(\lambda)$ it acquires vertical sections, forming a mushroom-type profile in phase $3$. The transition $2\rightarrow 3$ is continuous, but its expressions can be seen in the mechanical properties later. Upon further increasing $\varepsilon_x$, the mushroom's cap widens, which strengthens the vdW interactions of the ripple with itself, until at $\varepsilon_x=\varepsilon_x^{34}(\lambda)$ they become strong enough to make transition to phase $4$ with collapsed ripples. The collapsed phase has many energy minima and the relevant geometries depend on the construction process of the cardboard.\cite{Paronyan2011,kim_PRB_11,ortolani_NL_12} Therefore, it is sensible only to estimate the threshold for the collapse (see Supplemental Material \cite{supplemental}); the collapsed phase itself is not interesting because of properties similar to plain multilayer graphene.




The above-mentioned scale $\lambda_{co}$ determines the crossover from bending- to vdW-dominated energetics. Namely, the onset of rippling arises at small $\lambda$ from the competition between strain and bending energies, and at large $\lambda$ from the competition between strain and vdW energies. Therefore the crossover scale $\lambda_{co}$ occurs when the bending and vdW energies are equal, implying $\kappa A_\text{min}^2 \pi^2/\lambda_\text{co}^4=15 \varepsilon_\text{vdW} A_\text{min}^2/h^2$, or $\lambda_\text{co} \approx 10$~\AA. Above $\lambda_{co}$ graphene cardboard's properties evolve monotonously.

A closer investigation of the energy contributions reveals clear trends (Fig.\,\ref{fig:phases}b). The maximum strain in the middle layer prior to rippling for $\lambda>\lambda_{co}$ is $\varepsilon_{x,max}=\varepsilon_x^{12}$, corresponding to strain energy $21$~meV/\AA$^2$. After rippling, however, the strain energy vanishes and the elastic energy becomes bending-dominated with $E_\text{bend}\sim \lambda^{-2}$. Because this energy decreases rapidly upon increasing $\lambda$, bending and strain are comparable only at $\lambda$ close to or smaller than $\lambda_{co}$. The vdW energy decreases monotonously upon increasing $\varepsilon_x$. In principle rippling also creates compressive stress in $x$-direction, but in practice the induced strains are below $\sim 0.1$~\%\ at $\lambda_\text{co}$ and decrease rapidly with increasing wavelength.  

The cardboard can be further characterized by certain interesing properties. First, the maximum curvature within the ripples is $K_\text{max} \sim 0.3$~\AA$^{-1}$, a universal curvature of an exfoliating graphene-graphene interface. An estimate for this scale is obtained by equating the competing bending and vdW energies as $K_\text{max}\approx \sqrt{2\varepsilon_\text{vdW}/\kappa}= 0.2$~\AA$^{-1}$. Second, an estimate for the ratio $A/\lambda=2/\pi \sqrt{\varepsilon_x/(1-\varepsilon_x)}$ from Eq.(\ref{eq:sine}) is unexpectedly close to the observed ratio even when ripples deviate from the sine-profile. Third, material porosity ranges from zero at small $\varepsilon_x$ up to $90$~\%\ at $\lambda=100$~\AA\ and $99$~\%\ at $\lambda = 1$ $\mu$m; the maximum porosity is limited due to the collapsing instability.

\begin{figure}
\includegraphics[width=8cm]{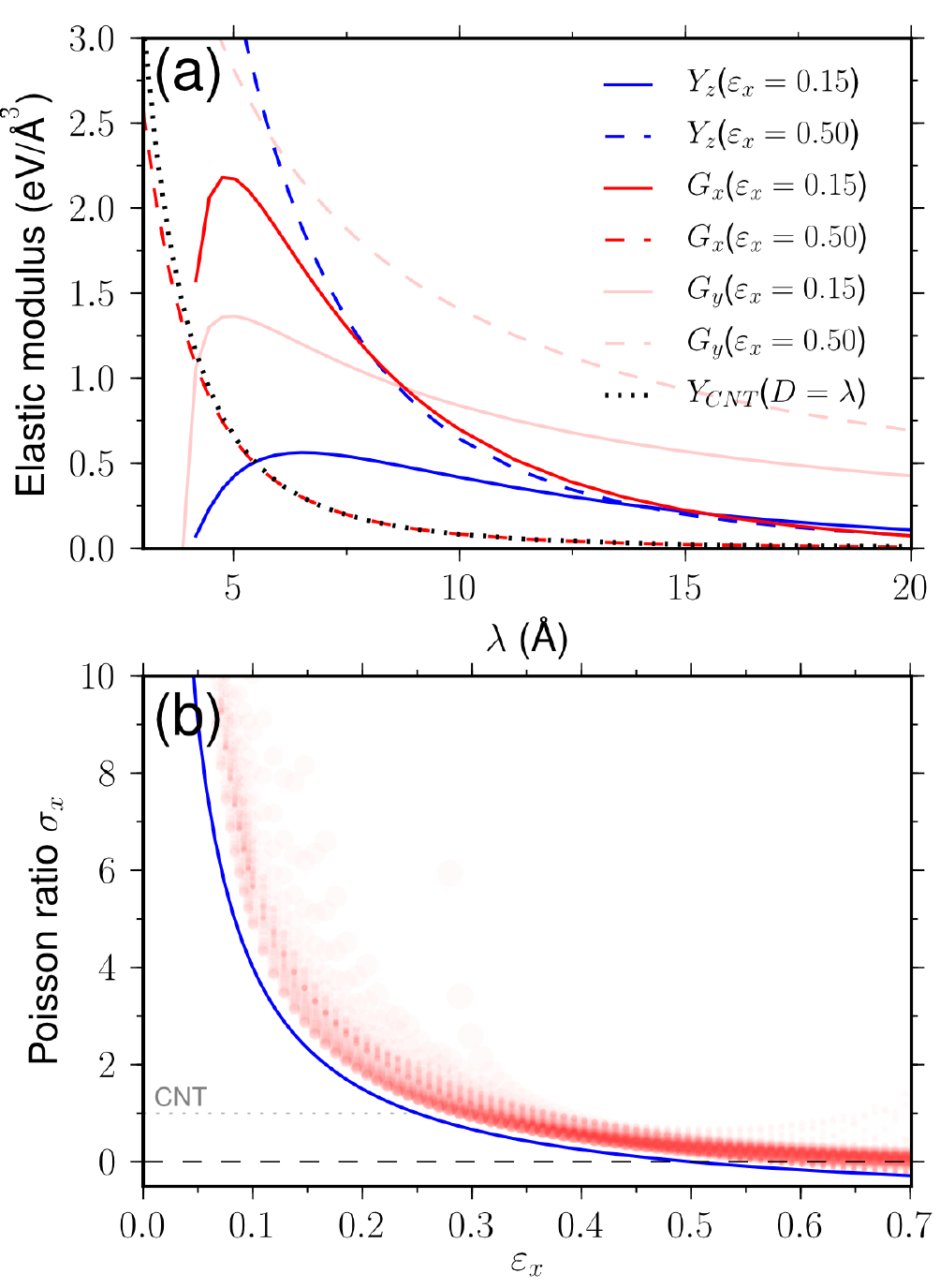}
\caption{The elastic parameters of graphene cardboard. (a) Young's modulus ($Y_z$) and shear moduli ($G_x$ and $G_y$) for different $\varepsilon_x$. Comparison is made with Young's modulus for a bundle of single-walled carbon nanotubes with diameters ($D=\lambda$). (b) Poisson ratio $\sigma_{xz}=-\Delta A/\delta \lambda$ superimposed for all $\varepsilon_x$ (transparent circles with size proportional to $\lambda$). Line is an analytical estimate, Eq.(\ref{eq:poisson}).}
\label{fig:elasticity}
\end{figure}

Particularly interesting are cardboard's mechanical responses to structural deformations (Fig.\,\ref{fig:elasticity}). The relevant deformations here are compression in $z$-direction, shear along $x$-direction, shear along $y$-direction, and tension along $x$-direction; other deformations are not as interesting regarding the relevant properties of the structure. Young's modulus for $z$-compression is dominated by ripples' bending energy, causing the modulus to decrease rapidly with increasing $\lambda$ (Fig.\,\ref{fig:elasticity}a). Thus, the modulus can be tuned greatly by varying $\lambda$. For comparison, the Young's modulus of a bundle of single-walled carbon nanotubes with a diameter of $D=\lambda$ is approximately $Y_\text{CNT}=18\pi \kappa/D^3$, which is close $Y_z(\lambda=D)$ with $\varepsilon_x=0.5$.\cite{moduli-normalization} This coincidence is plausible because at $\varepsilon_x\approx 0.5$ the ratio $A/\lambda$ is not far from one, meaning that the curvatures in the ripple and in the cylinder are similar. The modulus for shear along the ripples involves shear in the plane of the rippled graphene and can be solved as $G_y=4G_\text{gr}(1-\varepsilon_x)(A/\lambda)^2=16G_{gr}\varepsilon_x/\pi^2$, where $G_\text{gr}=9.5$~eV/\AA$^2$ is the graphene shear modulus.\cite{kudin_PRB_01} The shear modulus $G_x$, in turn, involves changes in $E_\text{bend}$ and decreases rapidly, roughly as $G_x \propto \lambda^{-2}$. Thus, the shear moduli of graphene cardboard are mostly anisotropic, but the degree and the sense of directionality depends both on $\lambda$ and $\varepsilon_x$.



Regarding Poisson ratios, it is opportune to note that the energy scale for strain in $x$-direction is much larger than the energy scale for strain in $z$-direction. For this reason a compression in $z$-direction does not change $\lambda$, rendering the Poisson ratio $\sigma_{zx}=0$. For the same reason, on the contrary, a tension in $x$-direction can change the ripple amplitude, implying a non-zero Poisson ratio $\sigma_{xz}=-\partial A/\partial \lambda|_{l=const.}$. Similarly, also $\sigma_{yz}$ is non-zero, but tension in $y$-direction only stretches graphene, which is not interesting in this context. While the Young's modulus for the strain in $x$-direction is simply $Y_x\approx 2k/A$, the Poisson ratio $\sigma_{xz}$ reveals far more interesting behavior (Fig.\,\ref{fig:elasticity}b). It shows \emph{qualitative tunability}, with values ranging from the maximum positive value of $\approx 10$ to a minimum negative value of $\approx -0.5$. While it also depends on wavelength, especially at small length scales, the main trend is captured by the dependence on $\varepsilon_x$. An estimate for the ratio from Eq.\,(\ref{eq:sine}) yields 
\begin{equation}
\sigma_{xz}=\frac{1}{2\varepsilon_x}-1, 
\label{eq:poisson}
\end{equation}
in good agreement with the general trend (Fig.\,\ref{fig:elasticity}b). The maximum of $\sigma_{xz}$ at the minimum accessible $\varepsilon_x=\varepsilon_x^{12}$ thus corresponds to ripples with amplitude $A_\text{min}$. The Poisson ratio decreases with increasing $\varepsilon_x$ until around $\varepsilon_x=0.25$ it corresponds to the ideal ratio of carbon nanotubes, which equals one independent of tube diameter. When $\varepsilon_x$ approaches $0.5$, the Poisson ratio goes to zero. This region corresponds to ripples close to the transition from sine- to mushroom-type profiles. Indeed, it is intuitively plausible that upon decreasing $\lambda$ (while keeping the length $l$ constant) sine-type profiles increase in height, mushroom-type profiles decrease in height, and profiles near the phase boundary $\varepsilon_x^{23}$ show heights that remain unchanged.

Although the model has its limitations, it should capture the main trends correctly. First, the choice for the distance between the ripple crest and the sandwitch has an effect only at small $\lambda$. Second, the choice for the form of vdW interaction, Eq.\,(1) in Supplementary Material, also has an effect only at small $\varepsilon_x$. Third, patchy covalent bonds between the ripple and the sandwitch layers would create sp$^3$-hybridized atoms and affect graphene bending modulus or create kinks. If the density such atoms is small, however, the importance of this effect should remain limited. Fourth, the model assumes a free-standing cardboard and ignores the role of substrates. Yet, it has been shown that dispersion forces in layered materials the are relevant solely for neighboring layers, at least in the absence of charging effects.\cite{bjorkman_PRL_12} Therefore, while stabilizing the bottom sandwitch layer, substrates would induce only minor quantitative displacements in the phase transition lines of Fig.\,\ref{fig:phases}a (principally only in $\varepsilon_x^{12}$); their presence would not change the overall picture. Fifth, even though in practice the sandwitch layers too will corrugate, its magnitude should remain small in view of the large energy invested in in-plane strain as compared to bending energy.\cite{Yamamoto2012a,pierre-louis_PRE_08}  
Extension of the model to multilayer ripples using revised $\kappa$ and $k$\cite{koskinen_PRB_10b} show similar phases, only length scales are larger; this extension is however limited to thin sheets due to delamination issues.\cite{Koskinen2013a}



The practical realization of graphene cardboard is undoubtedly a challenging problem, but there are different ways to approach it. One way is to create rippling by shear\cite{min_APL_11}, by mechanical loading\cite{neek-amal_PRB_10}, or by thermal ripple generation.\cite{Bao2009} Another way is to apply bending or pre-strain to flexible elastomers used as graphene substrates.\cite{poncharal_science_99,Sun2007} Even substrates themselves can be introduced with ripple-generating reconstructions or nanotrenches.\cite{tapaszto_nphys_12} The sandwitching of the ripples  with current experimental specifications is admittedly challenging, but perhaps not insurmountable.

To conclude, graphene cardboard has revealed mechanical characteristics that are tunable with respect to the phase-space parameters $\lambda$ and $\varepsilon_x$. The phase diagram thus obtained can serve as a starting point for investigations of electronic structures modified by the periodic modulations arising from bending and periodic contacts to the sandwhich layers. The experimental realization of the sandwitching process remains an open question, especially at small ripple wavelengths, but the trends shown here should provide a useful guideline and motivation to make the experimental efforts worthwhile.


I acknowledge the Academy of Finland for funding (grant number $251216$) and the Finnish IT Center for Science (CSC) for computational resources.


%

\clearpage

{\bf Supplemental Material}

Density-functional simulations were done by GPAW code\cite{mortensen_PRB_05,enkovaara_JPCM_10}, using a self-consistent vdW-DF xc-functional.\cite{Dion2004, Wellendorff2012} This functional describes reasonably the energetics and bonding in van der Waals solids, graphene in particular.\cite{reguzzoni_PRB_12} The simulation cell was periodic along $x$ and $y$ axes ($L_x=\lambda$, $L_y=2.46$~\AA), and non-periodic along $z$ axis (using $6.0$~\AA\ vacuum). Grid spacing was $0.20$~\AA\ and the reciprocal space had $4\times 10$ points. Ripple crests had a $h=3.4$~\AA\ distance constraint from the sandwitch layers and the optimizations used either constrained or unconstrained total thickness $H=A+2h$.\cite{bitzek_PRL_06}

The elastic model contains bending, stretching, and van der Waals contributions.\cite{landau_lifshitz} First, the bending energy is 
$
E_\text{bend} = \kappa/(2\lambda) \int K(l)^2 dl,
$
where $\kappa=1.47$~eV is the bending modulus,\cite{kudin_PRB_01} $K(l)=1/R$ is curvature with radius of curvature $R$, and integration is along the ripple. (Energies are per surface area of the cardboard.) Second, the strain energy is
$
E_s = k/(2\lambda) \int \epsilon(l)^2 dl,
$
where $k=22.0$~eV/\AA$^2$ is the in-plane modulus\cite{kudin_PRB_01} and $\epsilon(l)$ is the local strain. Third, regarding the van der Waals interaction, to a good approximation it can be considered only between adjacent layers.\cite{bjorkman_PRL_12} Thus, for the Lennard-Jones $12-6$ pair potential $V_{LJ}®=4\varepsilon_{LJ}[(h/r)^{12}-(h/r)^6]$, the vdW energy is 
\begin{equation}
E_\text{vdW} = \lambda^{-1}\int W_2(z(l)) dl,
\label{eq:evdW}
\end{equation}
where $W_2(z)$ is the vdW potential as seen by the rippled layer. The potential is $W_2(z)=W(z)+W(A-z)$, where $W(z-h)=-5/3 \varepsilon_\text{vdW} (h/z)^4[1-2/5(h/z)^6]$. The connection to Lennard-Jones parameters is $\varepsilon_\text{vdW}=6\pi \varepsilon_{LJ}h^2/(5A_c^2)$, where $A_c=2.62$~\AA$^{-1}$. The value $\varepsilon_{LJ}=3.0$~meV hence corresponds to $\varepsilon_\text{vdW}=18.3$~meV/\AA$^2$.\cite{Lebegue2013} 


When $E_\text{bend}$ dominates and $A/\lambda$ is small, $K\approx z''(x)$ and the ripple becomes $z(x)=A \sin^2 (x \pi/\lambda)$. This implies $A=2\lambda/\pi \sqrt{\varepsilon_x/(1-\varepsilon_x)}$, $E_{bend}=\kappa A^2\pi^4/\lambda^4$, and $E_\text{vdW}=15 \varepsilon_\text{vdW} A^2/h^2$. In the general case a parametrized surface $(x(t),z(t))$ was interpolated by cubic splines with $23$ points ($\times 11$ for integration); optimization constraints were as in DFT simulations.



The threshold for collapsing was estimated by equating the energies of a standing and a fully collapsed ripples. The energy of a ripple adhered within a lenght $L$ to other layers is $E=-\varepsilon_\text{vdW} L + E_\text{bend}$. A standing ripple has $L\approx \lambda$ and bending is dominated by four corners with $K\approx 0.1$~\AA$^{-1}$; a collapsed ripple has $L\approx \lambda[1+3/2\varepsilon_x/(1-\varepsilon_x)]$ and bending is dominated by two half-circles with $K\approx 2/h$ (depending on the precise geometry; \emph{cf.} Fig.\,3a). Thus the estimate for the instability threshold becomes
\begin{equation}
\varepsilon_x^{34}(\lambda)= 1- 1/[1+2(E_\text{bend}^\text{coll}-E_\text{bend}^\text{stand})/3\lambda \varepsilon_\text{vdW}],
\label{eq:instability}
\end{equation}
as shown in Fig.\,3a. 


\begin{thebibliography}{30}
\bibitem [{mod()}]{moduli-normalization}%
  \BibitemOpen
  \href@noop {} {}\bibinfo {note} {Both the Young's modulus and the shear
  moduli are normalized to the ripple volume $\propto A\lambda$.}\BibitemShut
  {Stop}%
\end{thebibliography}

%

\end{document}